\documentclass[12pt,a4paper]{article}
\usepackage{amsmath}
\usepackage{graphicx}
\pdfoutput=1
\usepackage{times}

\begin{document}
\begin{titlepage}
\title{Energy dependence of average transverse momentum in hadron production due to collective effects}
\author{ S.M. Troshin,  N.E. Tyurin\\[1ex] \small\it Institute
\small\it for High Energy Physics,\\\small\it Protvino, Moscow Region, 142281 Russia}
\date{}
\maketitle
\begin{abstract}
Motivated by the first
measurements of  the  experiment  CMS at the LHC at $\sqrt{s}=0.9$ and $2.36$ TeV, we discuss energy dependence 
of  average transverse momentum of the secondary particles in hadron production in pp collisions.
 We suggest a possible explanation of this dependence as a result of collective rotation of the transient state
and associate its further possible decrease with  flattening off at higher energies with transition 
to the genuine  QGP state of matter. \\[2ex]
\end{abstract}
\vfill
\end{titlepage}
\section*{Introduction}

Nowadays the LHC  enters the initial phase of collecting data and providing first experimental results. 
Along with realization of its discovery potential, start of the LHC experimental program will definitely lead
to the renewed interest to the well
known problems and deepened insights into those issues.
Multiparticle production and  global  observables such as average transverse momentum of
produced particles  provide  an information
on the mechanisms of deconfinement, hadronization as well as on the size and the temperature of an interacting system.
The first experimental results obtained at the initial LHC energies \cite{cmspt} $\sqrt{s}=0.9$ and $2.36$ TeV appear to be consistent
with the previous experimental measurements,  which demonstrate rising energy dependence of the average transverse
momentum of secondaries. They might imply description in the thermodynamics terms \cite{hag}. 
Origin of the energy dependent behavior can be related to the observation of the possible new states
of hadronic matter and therefore studies of this dependence  can 
provide clues on the phase transition between them as it was noticed by Van Hove in the context of the starting 
the $S\bar{p}pS$ experiments at CERN \cite{vhove}. It is interesting to note that the   experimentally observed
 one-particle transverse momentum
spectrum  with power-law tail is consistent with canonical Tsallis distribution associated with non-extensive thermodynamics 
(cf. e.g. \cite{biro}). Thus, the existing models relate energy dependence of transverse momentum with different mechanisms,
 but it is evident
that such dependence has a non-trivial origin and reflects some  gross features of the interaction dynamics.

In this note we suggest a new explanation of the energy dependence of the average transverse momentum
emphasizing that the collective effects in in the initial state of the 
 proton collisions might be a ground for this energy dependence.
Proceeding this way, we adhere to the similarities between hadron and nucleus collisions and use the geometrical notions for description
of multiparticle production in hadronic reactions  as it was done long ago by Chou and Yang \cite{chyn}. Of course, there should be 
obvious differences in hadron and nuclei cases related 
to the size of colliding systems, but those differences are mostly quantitative ones.

In the following section we discuss what kind of degrees of freedom are responsible for the transient state of matter in hadron interaction
and after that we use these notions to make conclusion on the nature the energy dependence of the average transverse momentum.  

\section{Transient state of matter in hadron interactions}
 It might happen  therefore that
the transient states of matter in hadron and nuclei collisions have the same nature   and originates from the
nonperturbative sector of QCD, which has degrees of freedom  associated with
the mechanism of spontaneous chiral symmetry breaking ($\chi$SB) in QCD \cite{bjorken}. Due to this mechanism 
transition of current into  constituent quarks occur, which are  the quasiparticles whose masses are comparable with  a typical
 hadron mass scale.
 These constituent quarks interact via exchange
of the Goldstone bosons. Goldstone bosons  are collective excitations of the condensate and are represented by pions (cf. e.g. \cite{diak}).
The  general form of the effective Lagrangian (${\cal{L}}_{QCD}\rightarrow {\cal{L}}_{eff}$)
 relevant for description of the non--perturbative phase of QCD
 includes the three terms \cite{gold} \[
{\cal{L}}_{eff}={\cal{L}}_\chi +{\cal{L}}_I+{\cal{L}}_C.\label{ef} \]
Here ${\cal{L}}_\chi $ is  responsible for the spontaneous
chiral symmetry breaking and turns on first.  To account for the
constituent quark interaction and confinement the terms ${\cal{L}}_I$
and ${\cal{L}}_C$ are introduced.  ${\cal{L}}_I$ and
${\cal{L}}_C$ do not affect the internal structure of the constituent
quarks.

The picture of a hadron consisting of constituent quarks embedded
 into quark condensate implies that overlapping and interaction of
peripheral clouds   occurs at the first stage of hadron interaction.
Nonlinear field couplings   transform then some part of the kinetic energy to
internal energy \cite{heis,carr}. 
As a result the massive
virtual quarks appear in the overlapping region and  the effective
field is generated that way. This field is generated by $\bar{Q}Q$ pairs and
pions strongly interacting with quarks. Pions themselves are bound states of constituent
quarks. During this stage the part of the effective Lagrangian ${\cal{L}}_C$ is turned off
(it turns on again in the final stage of the reaction) and  interaction is
 described by ${\cal{L}}_I$, its
possible form of ${\cal{L}}_I$ was discussed in \cite{diakp}.
The  effective field (transient phase) generation time $\Delta t_{eff}$
\[
\Delta t_{eff}\ll \Delta t_{int},
\]
where $\Delta t_{int}$ is the total interaction time. This assumption on the almost instantaneous
generation of the effective field has obtained support in the very short thermalization time revealed
in heavy-ion collisions at RHIC \cite{therm}.

 This picture assumes  deconfinement at the initial stage of
 the hadron collisions and  generation of common for both hadrons mean field during the first stage.
Such ideas were  used in the model \cite{csn} which has
been applied to description of elastic scattering. Massive virtual quarks play a role
of scatterers for the valence quarks in elastic scattering and
 their hadronization leads to
production of secondary particles in the central region. The mechanism of multiparticle production
in the central and fragmentation region was described in \cite{mult}.
 To estimate number
of such scatterers one could assume that certain  part of hadron energy carried by
the outer condensate clouds is being released in the overlap region
 to generate massive quarks. Then this number can be estimated  by:
 \begin{equation} \tilde{N}(s,b)\,\propto
\,\frac{(1-\langle k_Q\rangle)\sqrt{s}}{m_Q}\;D^{h_1}_c\otimes D^{h_2}_c
\equiv N_0(s)D_C(b),
\label{Nsbt}
\end{equation} where $m_Q$ -- constituent quark mass, $\langle k_Q\rangle $ --
average fraction of hadron  energy carried  by  the constituent valence quarks. Function $D^h_c$
describes condensate distribution inside the hadron $h$, and $b$ is an impact parameter of the colliding hadrons.

In the following we will concern particle production in the central region.
Since the quarks are constituent, it is natural to expect  direct proportionality between an average number  of secondary particles  and
number of virtual massive quarks appeared  in  collision of the  hadrons with a given impact parameter value \cite{mult}:
\begin{equation}\label{mmult}
\langle n  \rangle (s,b)=\alpha  N_0(s)D_C(b),
\end{equation}
where $\alpha$ is a  constant factor . The  geometrical picture of hadron collision discussed above
implies that at high energies and non-zero impact parameters
 the constituent quarks produced in overlap region carry  large orbital angular momentum.
 It  can be estimated
as follows
\begin{equation}\label{l}
 L(s,b) \propto  b \frac{\sqrt{s}}{2}D_C(b).
\end{equation}
Due to supposed strong interaction
between quarks this orbital angular momentum will lead to the coherent rotation
of the quark system located in the overlap region as a whole.
This rotation is similar to rotation of the liquid
where strong correlations between particles momenta exist \cite{qgpo}.
In what follows we argue that this collective coherent rotation would lead to the energy dependence of the
average transverse momentum and can explain experimentally observed rising behavior of this quantity.
\section{Average transverse momentum}
To calculate average transverse momenta we need to know single-particle inclusive cross-section.
We use unitarized expression for this quantity based on the rational representation
  for the scattering amplitude, recent discussion on this form of unitarization
can be found in \cite{dpe}.
The rational form of unitarization in quantum field theory
 is based on the relativistic generalization \cite{umat}
 of the Heitler equation of the relativistic damping theory \cite{heit}.

Unitarity condition for the elastic scattering amplitude $F(s,t)$
 can be written in the form
 \begin{equation}\label{un}
 \mbox{Im}F(s,t)=H_{el} (s,t)+H_{inel} (s,t),
 \end{equation}
 where $H_{el,inel}(s,t)$ are the corresponding elastic  and inelastic overlap function
 introduced by Van Hove \cite{vanh}.  The functions $H_{el,inel}(s,t)$ are related to the functions
$h_{el,inel}(s,b)$ and via the Fourier-Bessel transforms, i.e.
\begin{equation}\label{hel}
H_{el,inel} (s,t)=\frac{s}{\pi^2}\int_{0}^{\infty} bdb h_{el,inel}(s,b) J_0(b\sqrt{-t}).
\end{equation}
The elastic and inelastic cross--sections can be obtained as
follows:
\begin{equation}\label{selin}
\sigma_{el,inel}(s)\sim \frac{1}{s} H_{el,inel} (s,t=0).
\end{equation}

In the ($U$--matrix) approach the elastic scattering matrix in the impact
parameter representation has the form:
\begin{equation}
S(s,b)=\frac{1+iU(s,b)}{1-iU(s,b)}, \label{um}
\end{equation}
where $S(s,b)=1+2if(s,b)$ and $U(s,b)$ is the generalized reaction matrix, which is
considered to be an input dynamical quantity similar to the
eikonal function. Unitarity equation rewritten at high energies
for the elastic amplitude $f(s,b)$ has the form
\begin{equation}
\mbox{Im} f(s,b)=h_{el}(s,b)+ h_{inel}(s,b)\label{unt}
\end{equation}
where the inelastic overlap function
\[
h_{inel}(s,b)\equiv\frac{1}{4\pi}\frac{d\sigma_{inel}}{db^2}
\]
 is the sum of
all inelastic channel contributions.
Inelastic overlap function
is related to $U(s,b)$ according to Eqs. (\ref{um}) and (\ref{unt}) as follows
\begin{equation}
h_{inel}(s,b)=\frac{\mbox{Im} U(s,b)}{|1-iU(s,b)|^{2}}\label{uf},
\end{equation}
i.e.
\begin{equation}\label{sinel}
\sigma_{inel}(s)=8\pi\int_0^\infty bdb \frac{\mbox{Im} U(s,b)}{|1-iU(s,b)|^{2}}.
\end{equation}
It should be noted that
\begin{equation}\label{imu}
\mbox{Im} U(s,b)=\sum_{n\geq 3} \bar U_n(s,b),
\end{equation}
where $\bar U_n(s,b)$ is a Fourier--Bessel transform of the function
\begin{eqnarray}\label{unn}
\bar U_n(s,t) & = & \frac{1}{n!}\int \prod_{i=1}^n\frac{d^3q_i}{q_{i0}}\delta^{(4)}(\sum_{i=1}^n q_i-p_a-p_b)
U_n^*(q_1,....,q_n;p_{a}',p_{b}')\cdot\\
& &U_n(q_1,....,q_n;p_a,p_b)\nonumber.
\end{eqnarray}
Here the functions $U_n(q_1,....,q_n;p_a,p_b)$ and $U_n(q_1,....,q_n;p_{a'},p_{b'})$ correspond to the
ununitarized (input or ``Born'') amplitudes of the process
\[
a+b\to 1+....+n,
\]
and the process with the same final state and the initial state with different momenta $p_a'$ and $p_b'$
\[
a'+b'\to 1+....+n,
\]
respectively. They are the analogs of the elastic $U$-matrix
for the processes $2\to n$. The sum in the right hand side of the
Eq. (\ref{unn}) runs over all  inelastic final  states $|n\rangle$.
 
Then denoting via $\zeta$ a set of kinematical variables which characterizes
the kinematics of the final state particle $c$, 
the expression for the inclusive cross-section of the process $ab\to cX$ has the following form
\cite{tmf}:
\begin{equation}
\frac{d\sigma}{d\zeta}= 8\pi\int_0^\infty
bdb\frac{I(s,b,\zeta)}{|1-iU(s,b)|^2}\label{unp},
\end{equation}
where $I(s,b,\zeta)$ is the Fourier-Bessel transform of the functions which are defined
similar to Eq. (\ref{un}) but with the fixed kinematical variables $\zeta$ related to the particle $c$
in the final state.
It should be noted that the impact parameter $b$ is the variable conjugated to the variable $\sqrt{-t}$, where
$t=(p_a-p_a')^2$ and that the following sum rule is valid for the $I(s,b,\omega)$
\begin{equation}\label{sum}
\int d\zeta I(s,b,\zeta)=\langle n\rangle (s,b)\mbox{Im} U(s,b).
\end{equation}
The initial impact parameter $b$ is related to the impact parameters of the
secondary particles by relation \cite{sakai}
\begin{equation}\label{v}
\mathbf{b}=\sum_{i=1}^n x_i\mathbf{b}_i,
\end{equation}
where $x_i$ stands for the Feynman variable $x$ of the $i$-th particle.
Combining Eqs. (\ref{uf},\ref{unp},\ref{sum}), one can easily obtain the following relation
of the average transverse momentum $\langle p_\perp(s)\rangle$ with  $\langle p_T \rangle (s,b)$:
\begin{equation}\label{mpt}
 \langle p_T\rangle (s)=\frac{\int_0^\infty bdb \langle p_T\rangle (s,b) \langle n\rangle (s,b) h_{inel}(s,b)}
{\int_0^\infty bdb  \langle n\rangle (s,b)h_{inel}(s,b)}
\end{equation}
with inelastic overlap function $h_{inel}(s,b)$ given by Eq. (\ref{uf}).
\section{Coherent rotation of transient matter and energy dependence of average transvese momentum}
We are going now to evaluate energy dependence of the average transverse momentum and propose a possible
mechanism leading to this dependence. First of all it should be noted that
the  function $U(s,b)$  is represented in the model as a product of the
averaged quark amplitudes $\langle f_Q \rangle$,
\begin{equation} U(s,b) =
\prod^{N}_{i=1} \langle f_{Q_i}\rangle (s,b).\end{equation}  This factorization originates
from an assumption of a  quasi-independent  nature  of the valence
quark scattering, $N$ is the total number of valence quarks in the
colliding hadrons.  The essential point here
is the rise with energy of the number of the scatterers  like
$\sqrt{s}$. The $b$--dependence of the function
$\langle f_Q \rangle$  has a simple form
$\langle f_Q \rangle (s,b) \propto\exp(-{m_Qb}/{\xi} )$ which can be related to the formfactor of a constituent quark.
The resulting generalized reaction matrix $U$  gets
the following  form
\begin{equation} U(s,b) = g\left (1+\alpha
\frac{\sqrt{s}}{m_Q}\right)^N \exp(-\frac{Mb}{\xi} ), \label{x}
\end{equation} where $M =\sum^N_{Q=1}m_Q$.
Here $m_Q$ is the mass of constituent quark, which is taken to be
$0.35$ $GeV$. Other parameters have values obtained from the experimental data
fitting.  To make explicit calculations  we model for simplicity
the condensate distribution $D_C(b)$  by the exponential form.
Then the mean multiplicity in the impact parameter representation has the form
\begin{equation}\label{nsbex}
  \bar n (s,b)=\alpha N_0(s)\exp (-b/R_C)
\end{equation}
which resulting in the two-parametric power-like energy dependence
 of the experimentally measurable average multiplicity multiplicity
\[
 \langle n\rangle (s)=as^\delta\
 \]
This power-like dependence of mean multiplicity is in good agreement with the experimental data \cite{mult}. Note that
power index has the following relation with parameters of the model
\[
\delta={\frac{1}{2}\left(1-\frac{\xi }{m_QR_{C}}\right)}. 
\]

It should be noted that 
discovery of the deconfined state of matter has
been announced  by four major experiments at RHIC \cite{rhic}.
Despite the highest values of energy and density have been reached,
a genuine quark-gluon plasma QGP (gas of the free current quarks and gluons)
was not found. The deconfined state reveals the properties of the perfect liquid,
being a strongly interacting collective state and therefore it was labeled as sQGP.
Using similarity between hadronic  and nuclear interactions we assumed that transient state in
hadron interactions is also a liquid-like strongly interacting matter. It means as it was already 
mentioned in the beginning that the presence of large angular momentum in the overlap region
will lead to coherent rotation of quark-pion liquid. Of course, there should be experimentally
observed effects of this collective effect, one of them is the directed flow in hadron reactions,
with fixed impact parameter discussed in \cite{qgpo}. It is not impossible task to measure impact
parameter of collision in hadron reactions with the help of the event multiplicity studies. 
But effects averaged over impact parameter can be measured more easily using
standard experimental technics. So, it is natural to assume that the rotation of transient matter
will affect average transverse momentum of secondary hadrons in proton-proton collisions.
Let for beginning do not take into account  the other sources of the transverse momentum  and temporally
suppose that all average transverse momentum is a result of a coherent rotation of transient liquid-like state. 
Then the following relation can be invoked
\begin{equation} \label{ptl}
\langle p_T\rangle(s,b)=\kappa L(s,b),
\end{equation}
 where $L(s,b)$ is given by Eq. (\ref{l}) and $\kappa$ is a constant which has  dimension of inverse length.
It is natural to relate it with inverse hadron radius,  $\kappa\sim 1/R_h$.
Now calculating the integrals in Eq. (\ref{mpt}),
 we obtain the power-like dependence
of the average transverse momentum $\langle p_T \rangle (s)$ at high energies
\begin{equation}\label{apt}
\langle p_T \rangle (s) = cs^{\delta_C},
\end{equation}
where
\[
 \delta_{C}={\frac{1}{2}\left(1-\frac{\xi }{m_QR_{C}}\right)}.
\]
The value of  $\xi$ is fixed from the data on angular
 distributions \cite{csn} and for the mass of constituent quark
 the standard value $m_Q=0.35$ GeV is taken. Of course, besides collective
effects average transverse momentum would get contributions from other sources
such as thermal distribution proposed long time ago by Hagedorn \cite{hag}.
This part has no energy dependence and we take it into account by simple addition of the constant
term to the power-dependent one, i.e.:
\begin{equation}\label{apte}
\langle p_T \rangle (s) = a+cs^{\delta_C}
\end{equation}
Existing experimental data can be described well (cf. Fig.1) using Eq. (\ref{apte}) with parameters
$a=0.337$ GeV/c, $c=6.52\cdot 10^{-3}$ GeV/c and $\delta_C=0.207$.
\begin{figure}[h]
\begin{center}
\resizebox{8cm}{!}{\includegraphics*{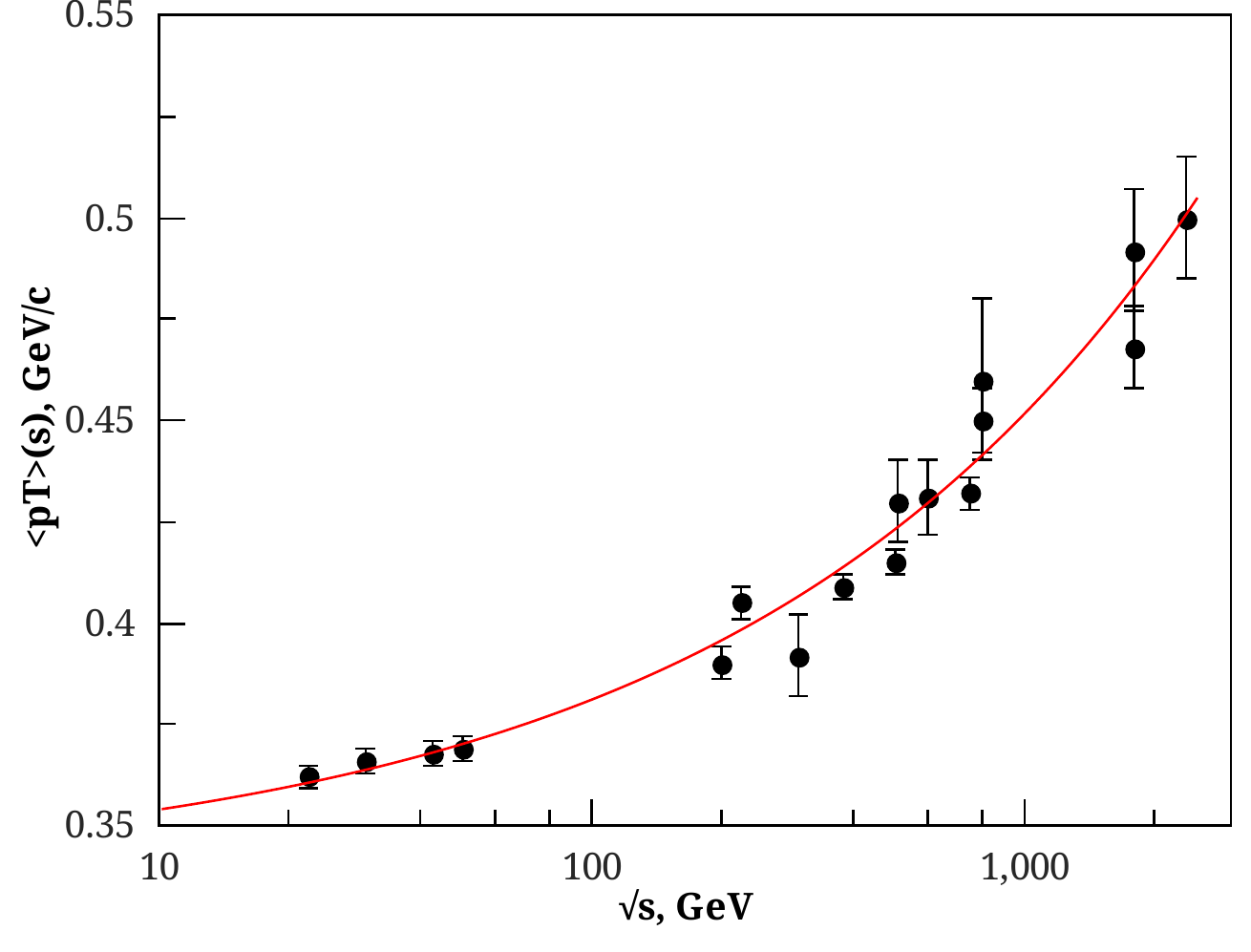}}
\end{center}
\caption{Energy dependence of the average transverse momentum in $pp$-collisions, experimental data from \cite{cmspt,all}.}
\end{figure}
The numerical value of $R_C$ is determined by pion mass with better than 10\% precision,
$ R_C\simeq 1/m_{\pi}$.
In the model the  indices in the energy dependencies of average multiplicity and transverse momentum $\delta$ and $\delta_C$ are
determined by the same expression and  experimental data fitting with free parameters $\delta$ and $\delta_C$ 
confirms this coincidence with better than 10\% precision also; note that the value $\delta=0.201$ follows from the experimental data analysis
for the average multiplicity. 
\begin{figure}[hbt]
\begin{center}
\resizebox{10cm}{!}{\includegraphics*{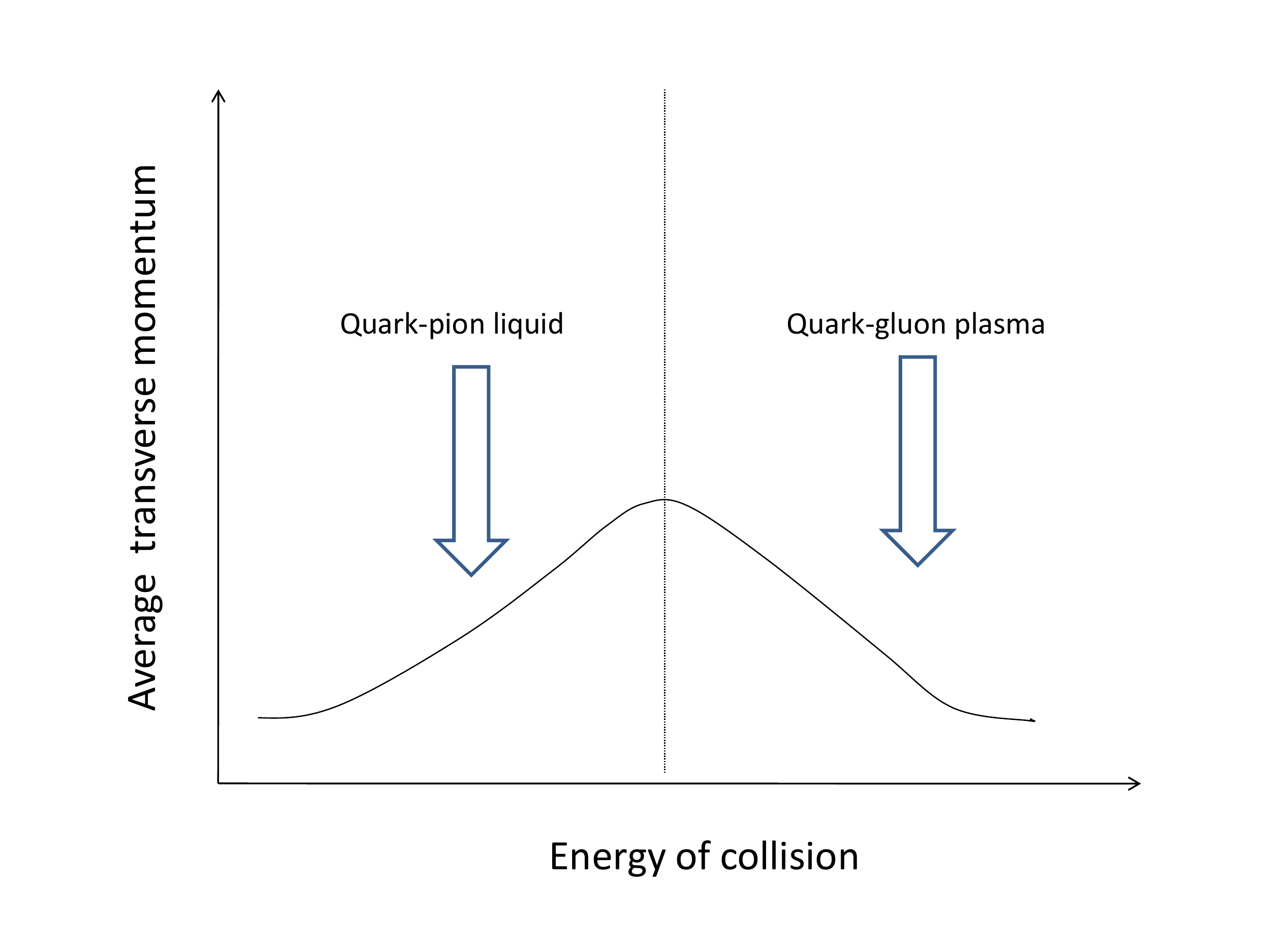}}
\end{center}
\vspace{-1cm}
\caption{Qualitative energy dependence of the average transverse momentum in $pp$-collisions in case of genuine QGP formation at
super high energies.}
\end{figure}
\section{Genuine QGP  formation}
It should be noted that the dynamical mechanism of the average transverse momentum growth originates in collective effect
of transient matter rotation, while dynamics of average multiplicity growth is related to the  mechanism where a nonlinear 
field couplings   transform  the kinetic energy to
internal energy. According to this difference one could expect divergent energy dependencies of these two observables at higher energies.
Indeed, formation of a genuine quark-gluon plasma in transient state in the form of the noninteracting gas of free quarks and gluons
would result in disappearance of the collective effect of rotation of the transient state. Orbital angular momentum presented 
in the initial state would lead then to the global polarization of produced particles as it was discussed in \cite{wang}. In \cite{qgpo} we noted
that disappearance of collective effect of rotation would lead  to the vanishing directed flow $v_1$. The simultaneous 
 vanishing of the energy dependent
contribution to the average transverse momentum should also be expected, i.e. average transverse momentum would reach maximum
at some energy and after that will decrease till eventual flat behavior (cf. Fig.2). Of course, this picture is a rather qualitative one since
the nature of phase transition from strongly interacting matter (associated with quark-pion liquid in the model) 
to the genuine quark-gluon plasma is not known. Currently, we  do  not even have any experimental indications
that such transition will indeed take place  at super high energies. It is interesting to note that inverse phase transition from parton
gas to liquid could explain a saturation phenomena in deep inelastic processes \cite{jenk}.
\section*{Conclusion}
We proposed here a new explanation of the energy dependent behavior of the average transverse momentum in pp collisions as
a result of the presence of a collective effect in the transient state of interaction. Due to strong interaction nature of this transient
state, the orbital angular momentum in the initial state would lead to collective rotation which in its turn contributes to average
transverse momentum. Orbital angular momentum increases with energy and this increasing behavior leads 
to increasing energy dependence of the average
transverse momentum. The main issue here is the assumption of the existence of the liquid intermediate  state in hadron interactions,
which is based on the analogy with nuclear collisions. Available experimental data are in good agreement with resulting model dependence. 
Transition to the genuine
QGP formation would destroy then any coherence (rotation) in the transient state and result in decreasing and further flattening of energy
dependence of the average transverse momentum. 

\end{document}